\pdfoutput=1

\documentclass[aps,pra,10pt,twocolumn,showpacs,superscriptaddress]{revtex4}
\usepackage{amsmath}
\usepackage{latexsym}
\usepackage{amssymb}
\usepackage{graphics,epstopdf}
\usepackage[colorlinks=true]{hyperref}

\begin{document}
\title{Degree of Complementarity Determines the Nonlocality in Quantum Mechanics }

\author{Manik Banik}
\email{manik11ju@gmail.com}
\affiliation{Physics and Applied Mathematics Unit, Indian Statistical Institute, 203 B.T. Road, Kolkata-700108, India}

\author{MD Rajjak Gazi}
\email{rajjakgazimath@gmail.com}
\affiliation{Physics and Applied Mathematics Unit, Indian Statistical Institute, 203 B.T. Road, Kolkata-700108, India}

\author{Sibasish Ghosh}
\email{sibasish@imsc.res.in}
\affiliation{Optics and Quantum Information Group, The Institute of Mathematical Sciences, C. I. T. Campus, Taramani, Chennai 600113, India}

\author{Guruprasad Kar}
\email{kar.guruprasad@gmail.com}
\affiliation{Physics and Applied Mathematics Unit, Indian Statistical Institute, 203 B.T. Road, Kolkata-700108, India}

\begin{abstract}
Bohr’s Complementarity principle is one of the central concepts in quantum mechanics which restricts joint measurement for certain observables. Of course, later development shows that joint measurement could be possible for such observables with the introduction of a certain degree of unsharpness or fuzziness in the measurement. In this paper, we show that the optimal degree of unsharpness, which guarantees the joint measurement of all possible pairs of dichotomic observables, determines the degree of nonlocality in quantum mechanics as well as in more general no-signaling theories.
\end{abstract}


\maketitle
At the time of the birth of quantum mechanics, two central non-classical concepts that appeared were the uncertainty principle \cite{heisen} and the complementarity principle \cite{bohr,bus,mitte,lahti}. Then came the most surprising idea of quantum nonlocality \cite{bell}. Recently it has been shown that quantum mechanics cannot be more nonlocal with measurements that respect the uncertainty principle \cite{oppen} (see also Ref. \cite{chefles}). In fact, this relationship between uncertainty and nonlocality holds well for all physical theories.

One of the original versions of the complementarity principle tells that there are observables in quantum mechanics that do not admit unambiguous joint measurement. Examples are position and momentum \cite{davice,prugo,bu}, spin measurement in different directions \cite{bu,kraus}, path and interference inn the double slit experiment \cite{woot, scully}, etc. With the introduction of the generalized measurement i.e. positive operator-valued measure (POVM), it was shown that observables which do not admit perfect joint measurement, may allow joint measurement if the measurements are made sufficiently fuzzy. This feature has been extensively studied for the case of position and momentum \cite{davice,prugo} and also for spin observables \cite{busch}, \cite{carmeli}.

In this paper, we argue that the complementarity principle can also play a role in determining the nonlocality for quantum mechanics as well as for various no-signaling probabilistic theories.  In particular, we show that the optimal degree of unsharpness that guarantees joint measurement of all possible pairs of dichotomic observables (such as s pair of dichotomic observables would exist in all physical theories) can be considered as the degree of complementarity and show that it determines the degree of nonlocality of the theory. Next we consider joint measurement for all possible pairs of observables with binary outcomes (which appears in Bell's inequality) in quantum mechanics and find the optimal unsharpness necessary for their joint measurement. This optimal unsharpness thus found, completely determines the degree of Bell violation in quantum mechanics.

Consider a generalized probability theory \cite{barrett} where any state of the system is described by an element $\omega$ of $\Omega$, the convex state-space of the system. $\Omega$ may be considered as a convex subset of a real vector space. By convexity of the state-space $\Omega$ we mean that any probabilistic mixture of any two states $\omega_1, \omega_2\in \Omega$, will describe a physical state of the system. An observable $\verb"A"$ (with the corresponding outcome set $\{\verb"A"_{j}~ : j \in J\}$) is an affine map from $\Omega$ into the set of probability distributions on the outcome set. A measurement of an observable $\verb"A" \equiv \{\verb"A"_j~|~\sum_jp^{\omega}_{\verb"A"_j}=1~~\forall~\omega\in\Omega\}$, performed on the system, allows us to gain information about the state $\omega$ of the physical system. The measurement of $\verb"A"$ consists of various outcomes $\verb"A"_{j}$ with $p^{\omega}_{\verb"A"_j}$ being the probability of getting outcome $\verb"A"_j$, given the state $\omega$. Let $\Gamma$ be the set of all observables with two measurement outcomes ( $j = + 1, - 1$ ), say `yes'$(=+1)$ and `no'$(=-1)$. If $\verb"A"\in\Gamma$ is such a kind of two-outcome observable, then the average value of $\verb"A"$ on a state $\omega$ is given by
$$\langle \verb"A"\rangle_\omega = p^\omega_{\verb"A"_{yes}}-p^\omega_{\verb"A"_{no}}~~.\eqno(1)$$
Given a two-outcome observable $\verb"A"\equiv\{\verb"A"_{yes},\verb"A"_{no}|~p^{\omega}_{\verb"A"_{yes}}
+p^{\omega}_{\verb"A"_{no}}=1~~\forall~\omega\in\Omega\}$, we define a fuzzy or unsharp observable, again with binary outcomes $\verb"A"^{(\lambda)}\equiv\{\verb"A"^{(\lambda)}_{yes},\verb"A"^{(\lambda)}_{no}~|~p^{\omega}_{\verb"A"^{(\lambda)}_{yes}}
+p^{\omega}_{\verb"A"^{(\lambda)}_{no}}=1~~\forall~\omega\in\Omega\}$, with `unsharpness parameter' $\lambda\in(0,1]$, where $p^{\omega}_{\verb"A"^{(\lambda)}_{yes(no)}}$ is the probability of getting the outcome $\verb"A"^{(\lambda)}_{yes(no)}$ in the measurement of $\verb"A"^{(\lambda)}$ with the result `yes' (`no'). The  probabilities $p^{\omega}_{\verb"A"^{(\lambda)}_{yes(no)}}$ are smooth versions of the probabilities of their original counterparts in the following way;
$$p^\omega_{\verb"A"^{(\lambda)}_{yes}}=\left(\frac{1+\lambda}{2}\right)p^\omega_{\verb"A"_{yes}}+\left(\frac{1-\lambda}{2}\right)p^\omega_{\verb"A"_{no}},\eqno(2)$$
for all $\omega\in\Omega$.

We denote the set of all unsharp observables with binary outcomes for a given $\lambda$  by $\Gamma^{(\lambda)}$. For any $\verb"A"^{(\lambda)}\in\Gamma^{(\lambda)}$ the average value of $\verb"A"^{(\lambda)}$ on a given state $\omega\in\Omega$ can be calculated as
$$\langle \verb"A"^{(\lambda)}\rangle_\omega = p^\omega_{\verb"A"^{(\lambda)}_{yes}}-p^\omega_{\verb"A"^{(\lambda)}_{no}}= \lambda \langle \verb"A"\rangle_\omega~~.\eqno(3)$$

Given a state $\omega\in\Omega$ and two observables $\verb"A"_1\equiv\{\verb"A"_{1j}~|~\sum_jp^{\omega}_{\verb"A"_{1j}}=1~~\forall~\omega\in\Omega\}$ and $\verb"A"_2\equiv\{\verb"A"_{2k}~|~\sum_kp^{\omega}_{\verb"A"_{2k}}=1~~\forall~\omega\in\Omega\}$, we say that joint measurement of $\verb"A"_1$ and $\verb"A"_2$ exists if there exists a joint probability distribution $\{p^{\omega}_{\verb"A"_{1j},\verb"A"_{2k}}|\sum_{j,k}p^{\omega}_{\verb"A"_{1j},\verb"A"_{2k}}=1\}$ satisfying the following conditions
$$\sum_{k}p^{\omega}_{\verb"A"_{1j},\verb"A"_{2k}}=p^{\omega}_{\verb"A"_{1j}},~~\forall~j\eqno(4a)$$\vspace{-.6cm} $$\sum_{j}p^{\omega}_{\verb"A"_{1j},\verb"A"_{2k}}=p^{\omega}_{\verb"A"_{1k}},~~\forall~k\eqno(4b)$$
whatever be the choice of $\omega\in\Omega$. For our purpose, we will concentrate only on the existence of the joint measurement of two two-outcome observables $\verb"A"_1,\verb"A"_2\in\Gamma$.  Given a physical theory it is not justifiable to demand that joint measurement should exist for any pair of $\verb"A"_1,\verb"A"_2\in\Gamma$, although in the classical world, it is always possible to construct a joint measurement observable. On the other hand, there are certain observables in quantum theory which can not be jointly measured jointly.

But it may be possible that observables which are not jointly measurable in a theory, may admit joint measurement for their unsharp counterparts within that theory. For two given observables, the values of unsharp parameter that make joint measurement possible, depend on the observables. Let $\lambda_{opt}$ denotes the optimum (maximum) value of the unsharp parameter $\lambda$ that guarantees the existence of joint measurement for $all$ possible pairs of dichotomic observables $\verb"A"^{(\lambda)}_1,\verb"A"^{(\lambda)}_2\in\Gamma^{(\lambda)}$. $\lambda_{opt}$ can then be considered as a property of that particular theory. It is obvious from the definition that joint measurement must exists for any two $\verb"A"^{(\lambda)}_1,\verb"A"^{(\lambda)}_2\in\Gamma^{(\lambda)}$, where $\lambda\leq\lambda_{opt}$. The value of $\lambda_{opt}$ measures the degree of complementarity of the theory in the sense that as $\lambda_{opt}$ decreases, the corresponding theory has more complementarity. Of course, finding the value of $\lambda_{opt}$ for a theory will depend on the details of the mathematical structure of the theory.

Let us now consider the case of a composite system consisting of two subsystems with associated state spaces ${\Omega}_1$ and ${\Omega}_2$ respectively (in a no-signaling probabilistic theory). The state space of the composite system is defined to be ${\Omega}_1 \otimes {\Omega}_2$, which is again a convex subset of a real vector space, whereas, an observable $\verb"A"_{12}$ is an affine map (with outcome space $\{\verb"A"_{12}^{(j)}~ :~ j \in J\}$) from ${\Omega}_1 \otimes {\Omega}_2$ into the set of all probability distributions on the outcome space \cite{barrett}. Given any observable $\verb"A"_{1}$ for the first subsystem (with outcome space $\{\verb"A"_{1j} : j \in J_1\}$) and any observable $\verb"A"_{2}$ for the second subsystem (with outcome space $\{\verb"A"_{2k} : k \in J_2\}$), here, for our purpose, we consider only observables of the form $\verb"A"_{12} = \{\verb"A"_{12}^{(jk)} : p^{\eta}_{\verb"A"_{12}^{(jk)}} = p^{\eta}_{\verb"A"_{1j}, \verb"A"_{2k}}~ {\rm for}~ {\rm all}~ (j, k) \in J_1 \times J_2~ {\rm and}~ {\rm for}~ {\rm all}~ \eta \in {\Omega}_1 \otimes {\Omega}_2\}$ where $p^{\eta}_{\verb"A"_{1j}, \verb"A"_{2k}}$ is the probability of getting the result $(j, k)$ when measurement of $\verb"A"_{1}$ and $\verb"A"_{2}$ are performed on the joint state $\eta$. Thus, when we take the unsharp version $\verb"A"^{(\lambda)}$ of a dichotomic observable for the first subsystem and a dichotomic observable $\verb"B"$ for the second subsystem then, for any state $\eta \in {\Omega}_1 \otimes {\Omega}_2$, $p^{\eta}_{\verb"A"^{(\lambda)}_{yes}, \verb"B"_{yes}}$ will be the unsharp version of the probability $p^{\eta}_{\verb"A"_{yes}, \verb"B"_{yes}}$, {\it i.e.}, $p^{\eta}_{\verb"A"^{(\lambda)}_{yes}, \verb"B"_{yes}} = (1/2 + {\lambda}/2)p^{\eta}_{\verb"A"_{yes}, \verb"B"_{yes}} + (1/2 - {\lambda}/2)p^{\eta}_{\verb"A"_{no}, \verb"B"_{yes}}$, etc.

\textbf{Theorem ($1$)}: Consider a composite system composed of two subsystem with state spaces $\Omega_1$ and $\Omega_2$ respectively in a no-signaling probabilistic theory. For any pair of dichotomic observabless $\verb"A"_1,\verb"A"_2\in\Gamma_1$  on the first system and dichotomic observables $\verb"B"_1,\verb"B"_2\in\Gamma_2$  on the second system with the joint state $\eta\in\Omega_1 \otimes \Omega_2$, we have the following inequality;
$$|\langle \verb"A"_1\verb"B"_1\rangle_\eta+\langle \verb"A"_1\verb"B"_2\rangle_\eta+\langle \verb"A"_2\verb"B"_1\rangle_\eta-\langle \verb"A"_2\verb"B"_2\rangle_\eta|\leq\frac{2}{\lambda_{opt}}\eqno(5)$$
where $\lambda_{opt}$ has the meaning as described above.

This theorem easily follows from the following result by Andersson \emph{et al.} \cite{bar}: In a no-signaling theory, if we consider a pair of observables $\verb"A"_1,\verb"A"_2\in\Gamma_1$ on one system and $\verb"B"_1,\verb"B"_2\in\Gamma_2$ on another and make the further assumption that on one side joint measurement is possible,
 then the following inequality holds.
$$|\langle \verb"A"_1\verb"B"_1\rangle_\eta+\langle \verb"A"_1\verb"B"_2\rangle_\eta+\langle \verb"A"_2\verb"B"_1\rangle_\eta-\langle \verb"A"_2\verb"B"_2\rangle_\eta|\leq2.\eqno(6)$$
This inequality may not hold, in general, as joint measurement may not be possible as for example in quantum theory. As we discussed earlier, the joint measurement may still be possible in that case by taking the unsharp counterparts of $\verb"A"_1,\verb"A"_2$ (say). We assume that $\verb"A"^{(\lambda)}_1$ and $\verb"A"^{(\lambda)}_2$ are jointly measurable. Then from Eqn.(6) we get,
$$|\langle \verb"A"^{(\lambda)}_1\verb"B"_1\rangle_\eta+\langle \verb"A"^{(\lambda)}_1\verb"B"_2\rangle_\eta+\langle \verb"A"^{(\lambda)}_2\verb"B"_1\rangle_\eta-\langle \verb"A"^{(\lambda)}_2\verb"B"_2\rangle_\eta| \le 2,~ {\rm or}$$
$$|\langle \verb"A"_1\verb"B"_1\rangle_\eta+\langle \verb"A"_1\verb"B"_2\rangle_\eta+\langle \verb"A"_2\verb"B"_1\rangle_\eta-\langle \verb"A"_2\verb"B"_2\rangle_\eta|\leq \frac{2}{\lambda}\eqno(7)$$
where $\langle \verb"A"^{(\lambda)}\verb"B"\rangle_\eta = P^\eta_{\verb"A"^{(\lambda)}_{yes},\verb"B"_{yes}} - P^\eta_{\verb"A"^{(\lambda)}_{yes},\verb"B"_{no}}
-P^\eta_{\verb"A"^{(\lambda)}_{no},\verb"B"_{yes}} + P^\eta_{\verb"A"^{(\lambda)}_{no},\verb"B"_{no}}
= \lambda \langle \verb"A"\verb"B"\rangle_\eta$ etc.
If we set $\lambda=\lambda_{opt}$, the theorem then follows.

From the expression of inequality ($5$) it is clear that the amount of Bell violation is upper bounded by the unsharp parameter $\lambda_{opt}$, which is a characteristic of complementarity of that particular physical theory. As for example, in classical theory joint measurement of any two dichotomic observables is possible which means $\lambda_{opt}=1$. Contrary to this, we will show that, in quantum mechanics, the value of $\lambda_{opt}$ is $\frac{1}{\sqrt{2}}$. And, therefore, the amount of nonlocality of quantum theory respects the  Cirel'son bound $2\sqrt{2}$ \cite{cirel}.

In quantum mechanics any state of a $d$-dimensional system $S$ is described by a density operator acting on a $d$-dimensional Hilbert space $\mathcal{H}_S$ and
measurements are associated with POVMs. Consider the following two dichotomic POVMs  $\overline{\mathcal{M}}_j\equiv\{\mathcal{A}_j, \mathcal{I}_d-\mathcal{A}_j\}$ $(j=1,2)$ where $\textbf{0}\leq\mathcal{A}_j\leq\mathcal{I}_d$, $\mathcal{I}_d$ being the identity operator on the Hilbert space $\mathcal{H}_S$ and $\mathcal{A}_j$ acts on $\mathcal{H}_S$. The unsharp version of $\overline{\mathcal{M}}_j$ is of the form: $\overline{\mathcal{M}}^{(\lambda)}_j\equiv\{\mathcal{A}^{(\lambda)}_j=\frac{1+\lambda}{2}\mathcal{A}_j+\frac{1-\lambda}{2}(\mathcal{I}_d-\mathcal{A}_j), (\mathcal{I}_d-\mathcal{A}_j)^{(\lambda)}=\frac{1-\lambda}{2}\mathcal{A}_j+\frac{1+\lambda}{2}(\mathcal{I}_d-\mathcal{A}_j)\}$, here $0\le\lambda\leq1$. Now we will prove the following theorem about the joint measurability of unsharp versions of two dichotomic quantum observables.

\textbf{Theorem ($2$)}: Given any $d$-dimensional quantum system, joint measurement for unsharp versions of any two dichotomic observables $\overline{\mathcal{M}}_1$ and $\overline{\mathcal{M}}_2$ of the system is possible with the largest allowed value of the unsharpness parameter $\lambda_{opt}=\frac{1}{\sqrt{2}}$.

We will first briefly describe the condition of joint measurability of unsharp versions of two dichotomic projection valued measurements (PVM), when system Hilbert space $\mathcal{H}_S\cong\mathbb{C}^2$, introduced by Busch \cite{busch}.

Let $\mathcal{M}_j\equiv\{\wp_j=|\psi_j\rangle\langle\psi_j|,~ \mathcal{I}-\wp_j=|\psi^\bot_j\rangle\langle\psi^\bot_j|\}$ (for $j=1,2)$ be two dichotomic PVMs in $\mathbb{C}^2$ where $|\psi_j\rangle$, $|\psi^\bot_j\rangle$ are normalized pure states in $\mathbb{C}^2$ such that $\langle\psi_j|\psi^\bot_j\rangle=0$. The unsharp version of $\mathcal{M}_j$ be denoted as $\mathcal{M}^{(\lambda)}_j$. Joint measurement of $\mathcal{M}^{(\lambda)}_1 = \{\wp_1^{({\lambda})} \equiv ((1 + {\lambda})/2){\wp_1} + ((1 - {\lambda})/2)({\mathcal{I}}_2 - {\wp_1}), ({\mathcal{I}}_2 - {\wp_1})^{({\lambda})} \equiv ((1 - {\lambda})/2){\wp_1} + ((1 + {\lambda})/2)({\mathcal{I}}_2 - {\wp_1})\}$ and $\mathcal{M}^{(\lambda)}_2 = \{\wp_2^{({\lambda})} \equiv ((1 + {\lambda})/2){\wp_2} + ((1 - {\lambda})/2)({\mathcal{I}}_2 - {\wp_2}), ({\mathcal{I}}_2 - {\wp_2})^{({\lambda})} \equiv ((1 - {\lambda})/2){\wp_2} + ((1 + {\lambda})/2)({\mathcal{I}}_2 - {\wp_2})\}$ is possible iff there there exists a POVM $\mathcal{M}^{(\lambda)}_{12}\equiv\{G_{++},G_{+-},G_{-+},G_{--}\}$ such that each $G_{ij}$ is a positive operator on $\mathbb{C}^2$ ( for $i,j=+,-$ ) satisfying the following properties:
$$G_{++}+G_{+-}+G_{-+}+G_{--}=\mathcal{I}_2;~~~~~~~~~~~~~~~~~~~~~~~~~~~~\eqno(8)$$\vspace{-.7cm}
$$G_{++}+G_{+-}=\wp^{(\lambda)}_1~;G_{-+}+G_{--}=(\mathcal{I}_2-\wp_1)^{(\lambda)};~~~\eqno(9)\\$$\vspace{-.7cm}
$$G_{++}+G_{-+}=\wp_2^{(\lambda)}~;~G_{+-}+G_{--}=(\mathcal{I}_2-\wp_2)^{(\lambda)}.\eqno(10)\\$$
In the measurement of the POVM $\mathcal{M}^{(\lambda)}_{12}$ if $G_{++}$ `clicks', we say that both $\wp^{(\lambda)}_1$ as well as $\wp^{(\lambda)}_2$ have been `clicked', and so on. According to Busch \cite{busch}, the POVM $\mathcal{M}^{(\lambda)}_{12}$ will exist for all possible pairs of unsharp measurement iff $0< \lambda \leq \frac{1}{\sqrt{2}}$. Thus the maximum allowed value of the unsharpness parameter $\lambda_{opt}$ is $\frac{1}{\sqrt{2}}$. In the case of PVM, the factor $\frac{1+\lambda}{2}$ ($\frac{1-\lambda}{2}$) has been interpreted as degree of reality (unsharpness) \cite{busch}. Before describing the proof of Theorem (2), we prove the following Lemma.

\textbf{Lemma}: Joint measurement of unsharp versions of any two dichotomic PVMs $\mathcal{M}_1$ and $\mathcal{M}_2$ of any $d$-dim quantum system is possible iff $0< \lambda \leq \frac{1}{\sqrt{2}}$.

\textbf{Proof}: Let $\mathcal{M}_j$ ($j=1,2$) be two PVMs in $\mathcal{H}_S\cong\mathbb{C}^d$ with unsharp versions  $\mathcal{M}^{(\lambda)}_j\equiv\{\wp^{(\lambda)}_j=\frac{1+\lambda}{2}\wp_j+\frac{1-\lambda}{2}(\mathcal{I}_d-\wp_j), (\mathcal{I}_d-\wp_j)^{(\lambda)}=\frac{1-\lambda}{2}\wp_j+\frac{1+\lambda}{2}(\mathcal{I}_d-\wp_j)\}$.
Each $\wp_j$ is a $d_j$-dimensional projector on $\mathbb{C}^d$ ( with $1\leq d_j\leq d-1$ for $j=1,2$ ) and $\mathcal{I}_d$ is the identity operator on $\mathbb{C}^d$ \cite{note}. $\wp_j$, $\mathcal{I}_d-\wp_j$, for $j=1,2$ are four projectors on $\mathbb{C}^d$ satisfying the conditions $\wp_j+(\mathcal{I}_d-\wp_j)=\mathcal{I}_d$. Now we use a powerful Lemma reformulated by L. Masanes in \cite{masa}, which tells that  there exists an orthonormal basis (ONB) of $\mathcal{H}_S$ such that, with respect to this basis, we have $\mathcal{H}_S=\bigoplus^{d'}_{\alpha=1}\mathcal{H}_{\alpha}$, where dimensions of the subspaces $\mathcal{H}_{\alpha}$ are either one or two and each of the four projectors $\wp_1, \mathcal{I}_d-\wp_1, \wp_2,$and $\mathcal{I}_d-\wp_2$ can be block-diagonalized. Hence, $\wp_1=\sum^{d'}_{\alpha=1}\wp^{(\alpha)}_1$ where the projectors $\wp^{(\alpha)}_1$'s are orthogonal to each other, $\wp^{(\alpha)}_1$ is supported on $\mathcal{H}_{\alpha}$ and the rank of  $\wp^{(\alpha)}_1$ can be either $0$, $1$, or $2$. Similarly, it holds for the other three projectors.   Now, given any $\alpha\in\{1,2,...,d'\}$, one can have one and only one of the following cases: rank of $\wp^{(\alpha)}_1=i$ and $\wp^{(\alpha)}_2=j$ with $i,j=0,1,2$.\\
Consider the case when the rank of $\wp^{(\alpha)}_1=0=$ the rank of $\wp^{(\alpha)}_2$, thus we have $(\mathcal{I}_d-\wp_1)^{(\alpha)}=(\mathcal{I}_d-\wp_2)^{(\alpha)}=\mathcal{I}_{\mathcal{H}_\alpha}$. So $\mathcal{M}^{(\alpha,\lambda)}_1\equiv\{\wp^{(\alpha,\lambda)}_1=\frac{1+\lambda}{2}\wp^{(\alpha)}_1+\frac{1-\lambda}{2}(\mathcal{I}_d-\wp_1)^{(\alpha)}
=\frac{1-\lambda}{2}\mathcal{I}_{\mathcal{H}_{\alpha}}, (\mathcal{I}_d-\wp_1)^{(\alpha,\lambda)}=\frac{1-\lambda}{2}\wp^{(\alpha)}_1+\frac{1+\lambda}{2}(\mathcal{I}_d-\wp_1)^{(\alpha)}
=\frac{1+\lambda}{2}\mathcal{I}_{\mathcal{H}_{\alpha}}$\}, where $\mathcal{I}_{\mathcal{H}_{\alpha}}$ is the identity operator acting on $\mathcal{H}_{\alpha}$. Similarly $\mathcal{M}^{(\alpha,\lambda)}_2\equiv\{\wp^{(\alpha,\lambda)}_2=\frac{1-\lambda}{2}\mathcal{I}_{\mathcal{H}_{\alpha}}
, (\mathcal{I}_d-\wp_2)^{(\alpha,\lambda)}=\frac{1+\lambda}{2}\mathcal{I}_{\mathcal{H}_{\alpha}}\}$. In this case, it is evident that the set  $\mathcal{M}^{(\alpha,\lambda)}_{12} $ of four positive operators does exist satisfying Eqs. (8)-(10) for $\alpha$-th block for all $\lambda$ with $0<\lambda\le1$. A similar analysis shows that for all other cases except in one, $\mathcal{M}^{(\alpha,\lambda)}_{12}$ always exists for all $\lambda$ with $0<\lambda\le1$. The only non-trivial case occurs when, the rank of $\wp^{(\alpha)}_1=$ the rank of $\wp^{(\alpha)}_2=1$. Let us say $\wp^{(\alpha)}_1=|\chi^{\alpha}_1\rangle\langle\chi^{\alpha}_1|$ and $\wp^{(\alpha)}_2=|\chi^{\alpha}_2\rangle\langle\chi^{\alpha}_2|$ where $|\chi^{\alpha}_1\rangle, |\chi^{\alpha}_2\rangle \in \mathcal{H}_{\alpha}$. We can have the following three subcases:\\
$(i)~|\chi^{\alpha}_1\rangle=|\chi^{\alpha}_2\rangle$ ( up to a global phase factor),\\
$(ii)~\langle\chi^{\alpha}_1|\chi^{\alpha}_2\rangle = 0,$\\
$(iii)~0<|\langle\chi^{\alpha}_1|\chi^{\alpha}_2\rangle|<1.$\\
For subcases $(i)$ and $(ii)$, it can be shown as above that $\mathcal{M}^{(\alpha,\lambda)}_{12}$ exists for all $\lambda$ with $0<\lambda\le1$, and the only non-trivial subcase $(iii)$ boils down to the case of existence of unsharp joint measurement of two non-commuting observables $\mathcal{M}^{(\alpha)}_1\equiv\{\wp^{(\alpha)}_1,(\mathcal{I}_d-\wp_1)^{(\alpha)}\}$ and $\mathcal{M}^{(\alpha)}_2\equiv\{\wp^{(\alpha)}_2,(\mathcal{I}_d-\wp_2)^{(\alpha)}\}$ in $\mathcal{H}_\alpha\cong\mathbb{C}^2$. And we know, in this case $\lambda_{opt}=\frac{1}{\sqrt{2}}$ \cite{busch}.

Now we have to show that for $0 \leq \lambda \leq \frac{1}{\sqrt{2}}$, there exists a set $\mathcal{M}^{(\lambda)}_{12}$ of four positive operators $\{G_{++},G_{+-},G_{-+},G_{--}\}$ which satisfies Eqs. (8)- (10) for $\mathcal{M}^{(\lambda)}_1$ and $\mathcal{M}^{(\lambda)}_2$. We have already observed that for $0 \leq \lambda \leq \frac{1}{\sqrt{2}}$, $\{G^{(\alpha)}_{++},G^{(\alpha)}_{+-},G^{(\alpha)}_{-+},G^{(\alpha)}_{--}\}$ exists for each block $\alpha$. We define the positive operators $G_{jk}=\sum^{d'}_{\alpha=1}G^{(\alpha)}_{jk}$ for $j,k\in\{+,-\}$. Obviously these new four positive operators would satisfy the desired Eqs. (8) - (10) corresponding to the whole Hilbert space guaranteeing the existence of joint measurement of $\mathcal{M}^{(\lambda)}_1$ and $\mathcal{M}^{(\lambda)}_2$.

From above we see that ${\lambda}_{opt} \ge 1/{\sqrt{2}}$. Let us now consider joint measurement of the unsharp versions ${\mathcal M}_1^{(\lambda)} = \{\wp_{1}^{(\lambda)} \equiv (\frac{1 + \lambda}{2})|0{\rangle}{\langle}0| + (\frac{1 - \lambda}{2})\sum_{j = 1}^{d - 1} \j{\rangle}{\langle}j|,~ ({\mathcal I}_d - \wp_{1})^{(\lambda)} \equiv (\frac{1 - \lambda}{2})|0{\rangle}{\langle}0| + (\frac{1 + \lambda}{2})\sum_{j = 1}^{d - 1} |j{\rangle}{\langle}j|\}$, ${\mathcal M}_2^{(\lambda)} = \{\wp_{2}^{(\lambda)} \equiv (\frac{1 + \lambda}{2})|+{\rangle}{\langle}+| + (\frac{1 - {\lambda}}{2})|-{\rangle}{\langle}-| + (\frac{1 - \lambda}{2})\sum_{j = 2}^{d - 1} |j{\rangle}{\langle}j|,~ ({\mathcal I}_d - \wp_{2})^{(\lambda)} \equiv (\frac{1 - \lambda}{2})|0{\rangle}{\langle}0| + (\frac{1 + {\lambda}}{2})|-{\rangle}{\langle}-| + (\frac{1 + \lambda}{2})\sum_{j = 2}^{d - 1} |j{\rangle}{\langle}j|\}$ of the two PVMs ${\mathcal M}_1 = \{\wp_{1} \equiv |0{\rangle}{\langle}0|,~ ({\mathcal I}_d - \wp_{1}) \equiv \sum_{j = 1}^{d - 1} |j{\rangle}{\langle}j|\}$, ${\mathcal M}_2 = \{\wp_{2} \equiv |+{\rangle}{\langle}+|,~ ({\mathcal I}_d - \wp_{2}) \equiv |-{\rangle}{\langle}-| + \sum_{j = 2}^{d - 1} |j{\rangle}{\langle}j|\}$ respectively, where $|{\pm}{\rangle} = (1/{\sqrt{2}})(|0\rangle + |1\rangle)$. For the joint measurability of these two unsharp observables, it is easy to see (by restricting our attention to the two dim. subspace spanned by $|0\rangle$ and $|1\rangle$) here that ${\lambda}_{opt} = 1/{\sqrt{2}}$. So, for the general case also, we must have that ${\lambda}_{opt} = 1/{\sqrt{2}}$.
$~~~~~~~~~~~~~~~~~~~~~~~~~~~~~~~~~~~~~~~~~~~~~~~~~~~~~~~~~~~~~~~~~~~~\Box$

\textbf{Proof of the theorem ($2$)}: As both $\overline{\mathcal{M}}_1 = \{\mathcal{A}_1, \mathcal{I}_d - \mathcal{A}_1\}$ and $\overline{\mathcal{M}}_2 = \{\mathcal{A}_2, \mathcal{I}_d - \mathcal{A}_2\}$ are dichotomic POVMs, therefore (by Neumark's dialation theorem \cite{peresbook}), one can find out a two dim. ancilla system $A$ with the associated Hilbert space ${\cal H}_A$ being spanned by an ONB $\{|0{\rangle}_A, |1{\rangle}_A\}$, such that the measurements of $\overline{\mathcal{M}}_1$ and $\overline{\mathcal{M}}_2$ can be realized by two dichotomic PVMs $\mathcal{M}_1 = \{\wp_{11}, (\mathcal{I}_d - \wp_{11})\}$ and $\mathcal{M}_2 = \{\wp_{12}, (\mathcal{I}_d - \wp_{12})\}$  respectively on the joint Hilbert space ${\cal H}_S \otimes {\cal H}_A$ such that for any state $\rho$ of the system $S$, we have: $Tr[\mathcal{A}_{l}\rho]=Tr[\wp_{1l}(\rho\otimes|0\rangle_A\langle0|)]$ and $Tr[(\mathcal{I}_d-\mathcal{A}_{l})\rho]=Tr[(\mathcal{I}_{2d}-\wp_{1l})(\rho\otimes|0\rangle_A\langle0|)]$ for $l = 1, 2$.

Consider now the following two PVMs on $\mathcal{H}_{S+A}$:
$\mathcal{M}_{l}=\{\wp_{1l},(\mathcal{I}_{2d}-\wp_{1l})\}$ for $l=1,2$.
The unsharp versions of these two PVMs are $\mathcal{M}^{(\lambda)}_{l}=\{\wp^{(\lambda)}_{1l}=\frac{1+\lambda}{2}\wp_{1l}+\frac{1-\lambda}{2}(\mathcal{I}_{2d}-\wp_{1l}),
(\mathcal{I}_{2d}-\wp_{1l})^{(\lambda)}=\frac{1-\lambda}{2}\wp_{1l}+\frac{1+\lambda}{2}(\mathcal{I}_{2d}-\wp_{1l})\}$ (with the same value of $\lambda$ as was taken for the joint measurement of
$\overline{\mathcal{M}}^{(\lambda)}_{l}=\{\mathcal{A}^{(\lambda)}_{l},(\mathcal{I}_{d}-\mathcal{A}_{l})^{(\lambda)}\}$ ) for $l=1,2$.
We have seen earlier that joint measurement of $\mathcal{M}^{(\lambda)}_{1}$ and $\mathcal{M}^{(\lambda)}_{2}$ is always possible with a POVM
$\mathcal{M}^{(\lambda)}_{12}=\{G_{++},G_{+-},G_{-+},G_{--}\}$ on $\mathcal{H}_{S+A}$ provided $\lambda\leq\frac{1}{\sqrt{2}}$.
So we assume here that $\lambda\leq\frac{1}{\sqrt{2}}$.
Note that here: $\textbf{0}\leq G_{jk}\leq \mathcal{I}_{2d}$ for all $j,k\in\{+,-\}$ and $\sum_{j,k}G_{jk}=\mathcal{I}_{2d}$;
$$G_{++}+G_{+-}=\wp^{(\lambda)}_{11}~~;~~G_{-+}+G_{--}=(\mathcal{I}_{2d}-\wp_{11})^{(\lambda)}$$\vspace{-.8cm}
$$G_{++}+G_{-+}=\wp^{(\lambda)}_{12}~~;~~G_{+-}+G_{--}=(\mathcal{I}_{2d}-\wp_{12})^{(\lambda)}$$
Define: $\overline{G}_{jk}\equiv{}_{A}\langle0|G_{jk}|0\rangle_{A}$ for all $j,k\in\{+,-\}$.
Let $|\psi\rangle_{S}\in\mathcal{H_{S}}$. Then ${}_{S}\langle\psi|\overline{G}_{jk}|\psi\rangle_{S}={}_{S}\langle\psi|_{A}\langle 0|G_{jk}|0\rangle_{A}|\psi\rangle_{S}\geq0$
as $G_{jk}\geq 0$ for all $j,k\in\{+,-\}$.
Again,${}_{S}\langle\psi |(\mathcal{I}_{d}-\overline{G}_{jk})|\psi\rangle_{S}={}_{S}\langle\psi|_{A}\langle0|(\mathcal{I}_{2d}-G_{jk})|0\rangle_{A}|\psi\rangle_{S}\geq0$ as $G_{jk}\leq \mathcal{I}_{2d}$ for all $j,k\in\{+,-\}$.
Thus we see that $0\leq \overline{G}_{jk}\leq \mathcal{I}_{d}$ for all $j,k\in\{+,-\}$.
Now ${}_{S}\langle\psi|\sum_{j,k} \overline{G}_{jk}|\psi \rangle_{S}={}_{S}\langle\psi|_{A}\langle0|\sum_{j,k}G_{jk}|0\rangle_{A}|\psi\rangle_{s}=
{}_{S}\langle\psi|_{A}\langle0|\mathcal{I}_{2d}|0\rangle_{A}|\psi\rangle_{s}=1$ for all $\psi\in\mathcal{H}_{S}$.
This implies $\sum_{j,k=}\overline{G}_{jk}=\mathcal{I}_{d}$.
Finally, ${}_{S}\langle\psi|(\overline{G}_{++}+\overline{G}_{+-})|\psi\rangle_{S}={}_{S}\langle\psi|_{A}\langle0|(G_{++}+G_{+-})|0\rangle_{A}|\psi\rangle_{S}
=Tr[\wp^{(\lambda)}_{11}(|\psi\rangle_{S}\langle\psi|\otimes|0\rangle_{A}\langle0|]=Tr[\mathcal{A}^{(\lambda)}_{1}|\psi\rangle_{S}\langle\psi|]$ for all
$|\psi\rangle_{S}\in\mathcal{H}_{S}$, etc. Therefore we have
$$ \overline{G}_{++}+\overline{G}_{+-}=\mathcal{A}^{(\lambda)}_{1}~;~\overline{G}_{-+}+\overline{G}_{--}=(\mathcal{I}_{d}-\mathcal{A}_{1})^{(\lambda)};$$
\vspace{-.8cm}
$$\overline{G}_{++}+\overline{G}_{-+}=\mathcal{A}^{(\lambda)}_{2}~;~\overline{G}_{+-}+G_{--}=(\mathcal{I}_{d}-\mathcal{A}_{2})^{(\lambda)}.$$
It follows that $\overline{\mathcal{M}}^{(\lambda)}_{12}\equiv\{\overline{G}_{++},\overline{G}_{+-},\overline{G}_{-+},\overline{G}_{--}\}$ is an observable on $\mathcal{H}_{S}$
corresponding to the joint measurement of the unsharp (dichotomic) POVMs $\overline{\mathcal{M}}^{(\lambda)}_{1}$ and $\overline{\mathcal{M}}^{(\lambda)}_{2}$, with $\lambda\leq \frac{1}{\sqrt{2}}$. It then follows from the Lemma that ${\lambda}_{opt} = 1/{\sqrt{2}}$.~~~~~~~~~~~~~~~~~~~~~~~~~~~~~~~~~~~~~~~~~~~~~~~~~~~~~~~~~~~~~$\Box$

It follows from Theorem (1) that when two dichotomic observables $\verb"A"_1$ and $\verb"A"_2$ are not jointly measurable, then possibility of the joint measurement of their unsharp versions sets a bound on the Bell-Clauser-Horne-Shimony-Holt (Bell-CHSH) expression (Eq. (7)) not at $2$ but at some value higher than $2$. This gives rise to the possibility of violation of the Bell-CHSH inequality involving these two observables. Wolf et al. \cite{wolf} show that this possibility turns out to be a reality for all pairs of incompatible dichotomic observables in quantum theory.

It is to be noted that the uncertainty relation (UR) in a no-signaling theory, by itself, does not determine the bound on non-locality. For example, there is no uncertainty in both the Popescu-Rohrlich (PR)-correlation as well as the classical world \cite{oppen}. But these two theories are on opposite poles in the context of non-locality. Actually the UR provides a meaningful bound on non-locality of a theory when non-locality arises due to some non-classical effect, namely steering. On the other hand, the assumption on complementarity in any no-signaling theory could derive a Bell's inequality which then sets a bound on non-locality. Possibly this happens because the UR is related to the statistical nature of the theory whereas complementarity is deeply related to the detailed structure of the theory. For example, for the PR-correlation, there is no structure to exploit to find ${\lambda}_{opt}$.

In this context, the comment made by Oppenheim and Wehner \cite{oppen} is worth mentioning. They have shown that there may exist a theory that can be as nonlocal as quantum mechanics but that has less complementarity which apparently contradicts our result. From our result it can be inferred that  the notion of complementarity, namely information complementarity, that has been considered in Ref. \cite{oppen}, is not equivalent to the one used in this paper. We think that the notion of complementarity in terms of the non-existence of joint measurement is somewhat stronger and our results suggest further exploration of relations among different notions of complementarity.

\textbf{Note}: Recently, we found a recent paper by Busch \emph{et al.} \cite{busch2012}, where it has been shown that when the unsharp parameters $(\lambda, \mu)$ for the pair of dichotomic quantum observables $(\verb"A"_1, \verb"A"_2)$ are different, the joint measurability condition would give rise to ${\lambda}^2 + {\mu}^2 \le 1$, a condition which follows from the Cirel'son bound \cite{cirel}  -- the bound, we are aiming to achieve in the present paper in the special case $\lambda = \mu$.

\textbf{Acknowledgement}: The authors gratefully acknowledge the suggestions given by P. Busch to modify an earlier version of the manuscript together with bringing to our notice the work in ref. \cite{carmeli}. S.G. thankfully acknowledges the visit to the Physics and Applied Mathematics Unit of the Indian Statistical Institute, Kolkata,  during which part of the work has been performed. G.K. acknowledges support from the DST Project No. SR/S2/PU-16/2007.
\


\begin{thebibliography}{99}
\bibitem{note}Note that if $d_j = 1$ for some $j \in \{1, 2\}$, then both $\mathcal{M}_1$ as well as $\mathcal{M}_j$ are jointly measurable, as in this case $\wp_j = {\mathcal{I}}_d$.
\end{thebibliography}
\end{document}